\newlength{\extralineskip}
\documentstyle[12pt]{article}
\begin{document}
\begin{titlepage}
\begin{flushright}
          \begin{minipage}[t]{12em}
          \large UAB--FT--394\\
                 June 1996
          \end{minipage}
\end{flushright}

\vspace{\fill}

\vspace{\fill}

\begin{center}
\baselineskip=2.5em

{\LARGE  MAJORANA NEUTRINOS \\
AND LONG RANGE FORCES}
\end{center}

\vspace{\fill}

\begin{center}
{\sc J.A. Grifols, E. Mass\'o, and R. Toldr\`a}\\

     Grup de F\'\i sica Te\`orica and Institut de F\'\i sica
     d'Altes Energies\\
     Universitat Aut\`onoma de Barcelona\\
     08193 Bellaterra, Barcelona, Spain
\end{center}

\vspace{\fill}

\begin{center}

\large ABSTRACT
\end{center}
\begin{center}
\begin{minipage}[t]{36em}
We establish that forces mediated by the exchange of a pair of
Majorana neutrinos differ from those due to Dirac neutrino
exchange.
\end{minipage}
\end{center}

\vspace{\fill}

\end{titlepage}

\clearpage

\addtolength{\baselineskip}{\extralineskip}
Two--neutrino--exchange mediates long range forces between macroscopic
objects. This has been recognized for quite a long time now
since the pioneering work by Feinberg and Sucher \cite{Feinberg},
who did the first calculation in the effective Fermi theory for the
weak interactions. The result has been rederived later, in the context
of the Standard Model to include the neutral current effects, by
the same authors \cite{Feinberg2} and by the authors 
in reference \cite{Hsu}. The
extremely tiny effect generated by the 2--neutrino 
force is, however, far
from the reach of actual experimental check. Indeed, compared
to their gravitational attraction, the force between two protons
1 cm apart is about $10^{-28}$ weaker. Due to the $r^{-6}$
behavior of this force, it is only at about $\sim 10\ \mbox{\AA}$ 
that the gravitational attraction felt by 
two protons equals in strength the
repulsion due to the neutrino mediated force. Clearly, at these 
distances, neither the gravitational nor the neutrino force can
compete with the Van der Waals type forces that provide for the 
cohesion of matter.

Recently, however, a claim has been raised by Fischbach that 
2--neutrino forces
could produce catastrophical consequences for compact objects such
as white dwarfs or neutron stars \cite{Fischbach}. Indeed, many 
body interactions
in dense media, far from being suppressed as compared to the 
conventional 2--body interactions, would lead to an unacceptably
large self--energy for a neutron star. Already 8--body interactions
would render a self--energy associated to 2--neutrino exchange 
forces that exceeds by many orders of magnitude the 
rest--mass--energy content of the star. Fischbach then proceeds
to argue that a mass in excess of 0.4 eV for the neutrino would
shorten the range of the interactions to an extent such that the
generated self--energy would be kept tolerably small. Of course,
should this be true, it has important implications both for
particle physics and for cosmology. However, these results
have been questioned in recent work by two different groups. In fact,
Smirnov and Vissani \cite{Smirnov} argue that low energy neutrinos
produced and subsequently captured in the core of
stars --a phenomenon described in 
ref. \cite{Smith-Loeb}-- fill a degenerate Fermi neutrino sea that 
blocks (Pauli effect) the free propagation of the low frequency 
neutrinos responsible for the long range force. The second group,
Abada {\it et al} \cite{Abada}, perform the self-energy calculation
by a different technique from Fischbach (essentially differing by
the way they sum up the N-body interaction effects), and 
reach the conclusion that it is negligible. From all
the above one realizes that the issue is far from being settled
and we expect new developments and insights to appear. Here, however,
we shall discuss a different aspect of neutrino mediated forces.

In the present work we address the question of long range forces
mediated by Majorana neutrinos. Actually, the above referred work
has dealt exclusively with Dirac neutrinos and one would like
to explore what happens with self--conjugate neutrinos. Majorana
neutrinos fit very naturally in modern particle physics scenarios
and it is because of this fact that we feel that a calculation
of the explicit form of the 2--neutrino force is relevant. After
all, the odds are high that neutrinos are really Majorana
particles. And if this is so, all consequences of their nature
should be thoroughly explored.

Since Dirac and Majorana neutrinos 
only differ in the low energy limit
and it is the exchange of low frequency neutrinos which is 
responsible for the large distance potential, we expect a 
qualitatively different behavior for Majorana mediated forces
(through the different symmetry character of the 2--fermion
wave--function associated to the indistinguishability of the
2--particle--intermediate state). 

The effective 4--fermion lagrangian density can be written --recall
that we are interested in the low momentum transfer static source
component of the 2--neutrino interaction-- as follows
\begin{equation}
{\cal L}_{int} = \frac{G_F}{2\sqrt 2} 
                 \ \bar{\nu}_M\gamma_5 \gamma_\mu \nu_M
           \ \bar{n} \gamma ^{\mu} \left( 1+g_A \gamma _5\right) n,
\end{equation}
specialized to the neutron case and where the Majorana neutrino
$\nu _M$ has a mass $m$\footnote{Of course the $m\neq 0$ requirement
is necessary if we want to distinguish physically between Majorana 
and Dirac neutrinos.}. What we essentially need is the Fourier
transform of the non relativistic amplitude shown in Fig. 1. Using
the dispersion theory approach of ref. \cite{Feinberg2} the
resulting potential can be cast in the form
\begin{equation} \label{potential}
V(r)= \frac{1}{4\pi^2 r} \int_{4m^2}^\infty dt \; abs M_t \; 
                                              e^{-r \sqrt t},
\end{equation} 
where $abs M_t$ stands for the absorptive part 
of the crossed amplitude
($t$--channel) in Fig. 1. It turns out to be 
\begin{equation}
abs M_t = \frac{G_F^2}{48\pi} t \left( 1-\frac{4m^2}{t} 
                                          \right)^{3/2} ,
\end{equation}
which displays the $\beta ^3$ behavior characteristic for a 
$P$--wave propagation. The integral in eq. (\ref{potential}) 
can be given in terms
of a modified Bessel function. The resulting potential reads
\begin{equation} \label{VMajorana}
V_M(r) = \frac{G_F^2 m^2}{8\pi^3 r^3} K_2(2mr) .
\end{equation}

For the sake of comparison, we quote the corresponding formulae
for a Dirac neutrino. We have
\begin{equation} 
abs M_t = \frac{G_F^2}{48\pi} t \left( 1-\frac{m^2}{t} \right) 
                \left( 1-\frac{4m^2}{t} \right)^{1/2} ,
\end{equation}
{\sl i.e.} a $S$--wave mode, and the potential (for non--zero
Dirac mass)
\begin{equation} \label{VDirac}
V_D(r) = \frac{G_F^2 m^3}{16\pi^3 r^2} K_3(2mr) , 
\end{equation}
equivalent to the expression given in \cite{Fischbach} (setting
$b=$1, where $b$ is defined in this reference).

Of course, both potentials coincide when $m=0$, since the distinction
between both types of neutrinos is superfluous in this case. They
both give
\begin{equation}
V(r) = \frac{G_F^2}{16\pi^3 r^5} ,
\end{equation}
which is the well--known result.

The asymptotic forms of $K_2$ and $K_3$ imply that for distances
much larger than the Compton wavelength of the neutrino, both
potentials exhibit qualitatively different behavior
\begin{eqnarray}
V_M(r) &=& \frac{G_F^2}{16} \left( \frac{m^3}{\pi^5 r^7} 
                \right)^{1/2} e^{-2mr} ,          \\
V_D(r) &=& \frac{G_F^2}{32} \left( \frac{m}{\pi r} \right)^{5/2}
		e^{-2mr} .
\end{eqnarray}

From equations (\ref{VMajorana}) and (\ref{VDirac}) we see that
for $m \neq 0$ (and $r \neq 0$) $V_M < V_D$ always. It is 
apparent that the same will hold true for the case of the
N--body  forces considered in \cite{Fischbach}. Therefore,
massive neutrinos, either Dirac or Majorana, will not 
generate an embarrassingly large self--energy provided
$m>0.4$ eV, which is the value derived by Fischbach by
requiring that no subvolume in the star contains a neutrino
exchange energy larger than its own mass.
This requirement, although sufficient
to prevent the energy problem to arise, might even not be
necessary, since the stellar energy crisis may well be 
fictitious as advocated by \cite{Smirnov,Abada}.
Be it one way or the
other, our purpose was quite independent of the self--energy
issue, i.e. to display the different character of Majorana
mediated long range forces.

We thank the Theoretical Astroparticle Network for support under
the EEC Contract No. CHRX-CT93-0120 ( Direction
Generale 12 COMA ). This work has been partially supported by
the CICYT Research Project Nos. AEN95-0815 and AEN95-0882.
R.T. acknowledges a FPI Grant from the Ministerio de
Educaci\'{o}n y Ciencia (Spain).

\newpage

\section*{Figure caption}
\bigskip
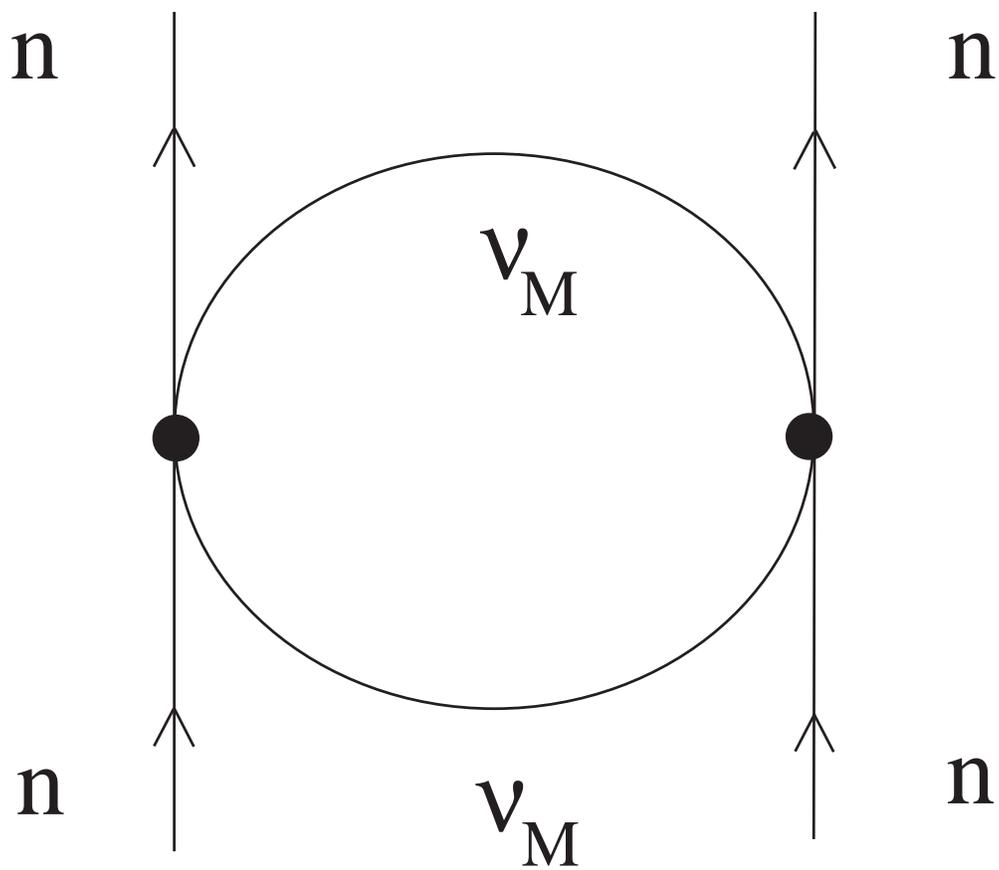
\noindent{\bf Figure 1:} Feynman diagram responsible for the
neutron-neutron potential arising from the exchange of 
two Majorana neutrinos.

\end{document}